\documentclass[twocolumn, prl, showpacs,preprintnumbers,amsmath,amssymb]{revtex4}

\usepackage{graphicx}
\usepackage{dcolumn}
\usepackage{bm}
\usepackage{amsmath}
\usepackage{latexsym}

\newcommand{\bra}[1]{\langle #1|}
\newcommand{\ket}[1]{|#1\rangle}

\newcommand {\be} {\begin{eqnarray}}
\newcommand {\ee} {\end{eqnarray}}
\newcommand{\ham}{\mathcal{H}}

\begin{document}

\preprint{APS/123-QED}

\title{Large nuclear Overhauser fields detected in vertically-coupled double quantum dots}

\author{Jonathan Baugh$^{1}$, Yosuke Kitamura$^{2}$, Keiji Ono$^{3}$ and Seigo Tarucha$^{2,4}$}
\affiliation{$^{1}$Institute for Quantum Computing, University of Waterloo, 200 University Ave. W., Waterloo, ON,  N2L 3G1}
\affiliation{$^{2}$Department of Applied Physics, School of Engineering, Tokyo University, 7-3-1 Hongo, Bunkyo-ku, Tokyo 113-8656, Japan}
\affiliation{$^{3}$Low Temperature Physics Laboratory, RIKEN, 2-1 Hirosawa, Wako, Saitama 351-0198, Japan}
\affiliation{$^{4}$ICORP-JST, 7-3-1 Hongo, Bunkyo-ku, Tokyo 113-8656, Japan}

\date{\today}

\begin{abstract}
We report the electrical induction and detection of dynamic nuclear polarization in the spin-blockade regime of double GaAs vertical quantum dots.  The nuclear Overhauser field is measured within a certain range, relying on bias voltage control of the interdot spin exchange coupling and measurement of dc current at variable external magnetic fields.  The largest Overhauser field observed was about 4 Tesla, corresponding to a nuclear polarization $\approx$ 40$\%$ for the electronic g-factor typical of these devices, $|g^*|\approx0.25$.  
The amount of polarization that can be detected using the present methods is limited by the range of externally applied magnetic fields over which a clear current step can be observed.  A phenomenological model based on the hyperfine mixing of electronic singlet and triplet spin states is proposed to explain these observations. 
\end{abstract}


\maketitle
Abundant nuclear spins present in semiconductor nanostructures couple collectively to the spins of confined electrons through the contact hyperfine interaction. In situations of mesoscopic confinement, an electron spin typically interacts with $N \sim 10^5 -10^6$ nuclear spins.  Gallium arsenide (GaAs) consists of $^{71}$Ga, $^{69}$Ga and $^{75}$As isotopes, each bearing nuclear spin $I=3/2$.  The hyperfine coupling constant in GaAs is $A\approx 90\mu$eV \cite{Paget}.  For completely polarized nuclear spins, this interaction leads to nuclear Overhauser fields $\sim$5-10 Tesla, depending on the precise value of the electron $g$-factor in the nanostructure \cite{Paget}.  Such large nuclear fields can produce measurable effects on spin-dependent electron transport; for example, in the spin-blockade (SB) regime of double quantum dots \cite{Ono, Ono_science} or in quantum well structures in the quantum Hall regime at special Landau level filling factors \cite{Hirayama}.  In the absence of near-perfect nuclear polarization, field fluctuations order $A/\sqrt{N}$ are the main source of dephasing for electron spin qubits in GaAs quantum dots \cite{Burkhard, Khaetskii, Merkulov, Petta}.  Dynamically polarizing the nuclear spin system either optically \cite{Imamoglu, Bracker} or electrically \cite{Rudner} has been proposed as one way to mitigate this effect \cite{Coish, Burkhard}.  It was also proposed that polarized ensemble of nuclear spins could serve as qubits or quantum memory, taking advantage of radio-frequency manipulation, transfer of information to electrons, and very long coherence times \cite{Taylor, Dobrovitski}.  Developing electrical control of the nuclear polarization in such devices would therefore be an important contribution to the present suite of experimental controls. \\
\indent Earlier experiments on transport in the SB regime of vertical coupled dots showed magnetic-field dependent current features such as a hysteretic current step and slow oscillations \cite{Ono}.  These features were identified with the presence of dynamic nuclear polarization (DNP), as well as complex feedback between the electronic and nuclear spin systems.  It has been proposed that DNP occurs in this system near a level degeneracy at which a blockaded spin-triplet state with one electron in each dot (e.g. $\ket{T_{-}}=\ket{\downarrow\downarrow}$) is mixed with the singlet state $\ket{S}=\frac{1}{\sqrt{2}}(\ket{\uparrow\downarrow}-\ket{\downarrow\uparrow})$ via the flip-flop terms of the hyperfine interaction, allowing transport through the dots \cite{Ono, TaruchaStatSol}.  A mutual electron/nuclear spin flip/flop occurs to conserve angular momentum,  and energy conservation is in general provided by inelastic processes \cite{Erlingsson}.  Since the hyperfine mixing is only efficient for one of the triplet states, the nuclear spins are pumped toward a polarized state.  Cotunneling processes due to strong coupling with the leads yield finite lifetimes for the triplet states not efficiently mixed by the hyperfine interactions. The accumulated nuclear polarization creates an Overhauser field (henceforth defined as the average Overhauser field of the two dots) that shifts the energy levels of the $\ket{T_{\pm}}$  triplet states.  In this Letter,  we determine the Overhauser field based on dc current measurement and voltage control of the interdot exchange coupling, with results that imply large nuclear polarizations.  The degree of nuclear polarization can be controlled by varying the external magnetic field. Finally, a phenomenological model is proposed to provide a qualitative understanding of these results. \\
\indent Experiments were performed on a $0.4 \mu$m diameter vertical double dot structure \cite{Tarucha, TaruchaStatSol, TaruchaSpinReview} with $10$ nm GaAs quantum wells, $7$ nm Al$_{0.3}$Ga$_{0.7}$As outer tunnel barriers and $6.5$ nm center barrier, at a sample temperature 1.7 K.  Gate voltage was adjusted to give a clear SB region in the current-voltage characteristic \cite{Ono, TaruchaStatSol}.  Measurements consisted of recording dc current as a function of source-drain voltage at variable external magnetic fields.  Crucially, the external field was applied in the plane of the two-dimensional electron gas (i.e. the lateral plane), so that it had negligible effect on the dot electronic wavefunctions. \\
\begin{figure}
\scalebox{0.11}{\includegraphics{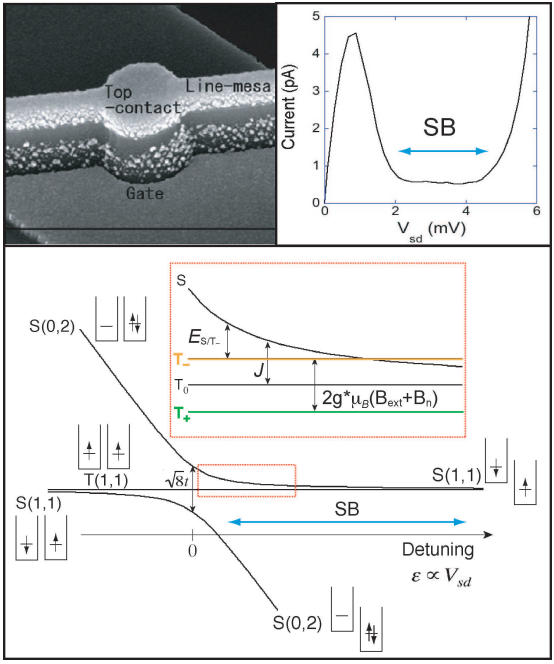}}
\caption{\label{fig:fig1} (upper, left) Scanning electron microscope image of a device nominally identical to the one investigated here. The double quantum dot is located in the pillar structure just below the surface of the gate metal. (upper, right)  Spin-blockade region (labeled `SB') in the I-V characteristic at zero external magnetic field ($B_{ext}=0$). (lower)  Schematic energy diagram of the two-electron eigenstate energies as a function of the relative detuning $\epsilon$ of the two dots.  The main figure shows the anti-crossing (resonance) of the (1,1) and (0,2) charge configurations of the spin singlet state, which are coupled by the interdot tunneling parameter $t$.  The (1,1) triplet state energy is independent of detuning. The expanded view shows the region just to the right of the resonance which is the focus of these experiments.  The Zeeman splitting of the $T_+/T_-$ states depends both on the external field $B_{ext}$ and the average nuclear Overhauser field $B_n$.  The spin exchange energy is defined as $J$, and the level separation between $\ket{T_-}$ and $\ket{S}$ is denoted by $E_{S/T_-}$. }
\end{figure}
\indent Figure~\ref{fig:fig1} shows a schematic energy level diagram for the two-electron states relevant to these experiments. The horizontal axis is the degree of energy detuning $\epsilon$ between the two dots, which is varied experimentally by changing the source-drain voltage $V_{sd}$.  In our device, $\epsilon\approx 0.27 V_{sd}+\epsilon_0$, with $\epsilon=0$ located slightly to positive $V_{sd}$, however the precise position is difficult to determine from analysis of the standard Coulomb diamond measurement.  The energy eigenvalues plotted in figure~\ref{fig:fig1} were calculated by taking into account intra- and interdot Coulomb energies, the interdot tunnel coupling $t$, and $\epsilon(V_{sd})$ \cite{TaruchaStatSol}. An interdot tunneling resonance between the (1,1) and (0,2) charge configurations of the spin singlet state is centered at $\epsilon=0$.  At large negative detuning, the (0,2) singlet is high in energy and Coulomb blockade occurs. Spin-blockade is mainly observed at positive detuning (i.e. to the right of the resonance).  The spin exchange coupling $J$ is defined as the energy difference between the singlet and spin-zero triplet $\ket{T_0}$, with a maximum value $J|_{\epsilon=0}=\sqrt{2}t$.  The expanded region in figure~\ref{fig:fig1} details the region of interest for our experiments. It shows that application of an external magnetic field $B_{ext}$ splits off the $\ket{T_{\pm}}$ states due to the Zeeman interaction.  Nuclear polarization arising in the two dots from the hyperfine mixing of $\ket{T_-}\rightarrow\ket{S}$ is antiparallel to $B_{ext}$, but causes an average Overhauser field $B_n\parallel B_{ext}$ since $sgn(gA)=-1$.  With respect to the $\ket{T_\pm}$ energies, $B_n$ and $B_{ext}$ are equivalent; this fact is central to our method for detecting $B_n$.  Note that the energy difference $E_{S/T_-}$ between $\ket{T_-}$ and $\ket{S}$ plays a leading role in the DNP process under study.  By varying detuning using $V_{sd}$, we can manipulate the field $B_{ext}+B_n$ for which $E_{S/T_-}=0$. \\
\begin{figure}
\scalebox{0.12}{\includegraphics{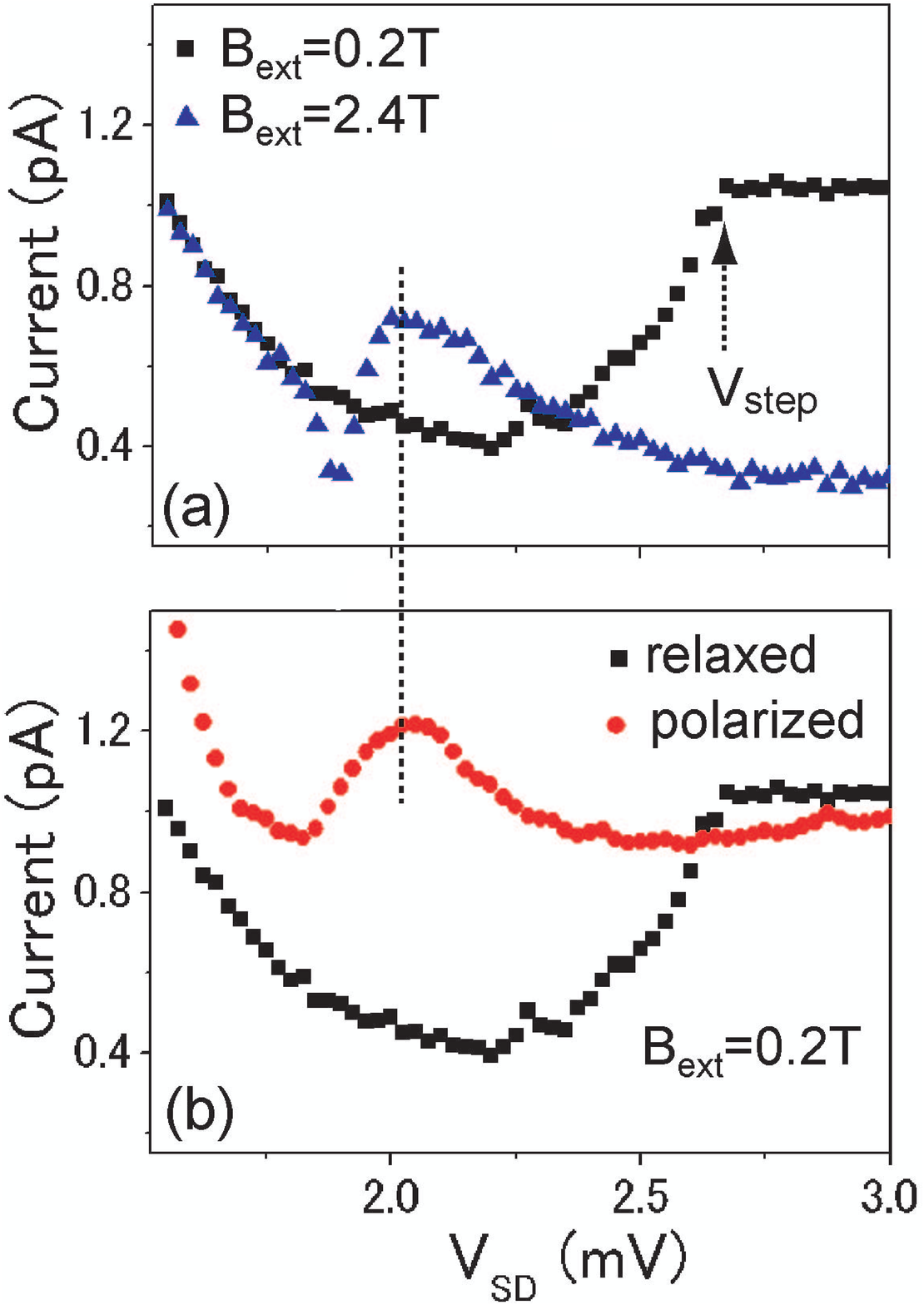}}
\caption{\label{fig:fig2} (a) Expanded view of the spin blockade region, showing current steps obtained by sweeping $V_{sd}$ (left to right) at $B_{ext}=0.2$ T and $B_{ext}=2.4$ T.  $V_{step}$ indicates the position of the 0.2 T step. A dot-lead resonance peak near $V_{sd}=1$ mV (see upper right of figure 1) gives rise to the background slope for $V_{sd}<2.0$ mV. (b) `Relaxed' trace is the $B_{ext}=0.2$ T trace in \textit{a}; `Polarized' trace was obtained at the same external field after pumping the nuclear polarization at $V_{sd}=3.0$ mV for 30s.}
\end{figure}
\indent Figure~\ref{fig:fig2}a shows that a current step of magnitude $\sim$0.5 pA  is observed when $V_{sd}$ is swept towards larger detuning at fixed external field $B_{ext}$ \cite{TaruchaStatSol}.  The characteristic step is very similar to that observed in Ref.~\cite{Ono} while sweeping $B_{ext}$ at fixed $V_{sd}$. This step appears to indicate the level crossing $E_{S/T_-}= 0$, and it shifts toward smaller $V_{sd}$ (i.e. smaller detuning) with increasing $B_{ext}$ as expected from figure~\ref{fig:fig1}.  We assign the step position $V_{step}$ to the edge of the upper current level, as shown in figure~\ref{fig:fig2}a.  No step was detected for $B_{ext}<0.1$T in the present device.  Traces were obtained at a sweep rate of 0.15 mV$s^{-1}$.\\
\indent  Figure ~\ref{fig:fig2}b shows that $V_{step}$ is also shifted in the presence of a nuclear Overhauser field.  To obtain the trace labeled `polarized', we first pumped the nuclear polarization by fixing $V_{sd}>V_{step}$ for the $V_{step}$ observed in the absence of an Overhauser field (trace labeled `relaxed').  After pumping about 30s, $V_{sd}$ is rapidly returned to $V_0=1$ mV and subsequently swept up again to obtain the second trace.  Both traces in figure~\ref{fig:fig2}b are taken at the same value of $B_{ext}=0.2$ T.  The position of the shifted current step is nearly identical to the $V_{step}$ obtained at larger $B_{ext}=2.4$ T in figure~\ref{fig:fig2}a.  From these experiments we conclude that the Overhauser field $B_{n}$ present in figure~\ref{fig:fig2}b can be estimated as the difference in external fields applied in figure~\ref{fig:fig2}a.  All such measurements were preceded by suitably long wait times at $V_{sd}=0$ to allow for relaxation of residual nuclear polarization from previous traces.  Relaxation of nuclear polarization occurring during the readout sweep is discussed below. \\
\begin{figure}
\scalebox{0.125}{\includegraphics{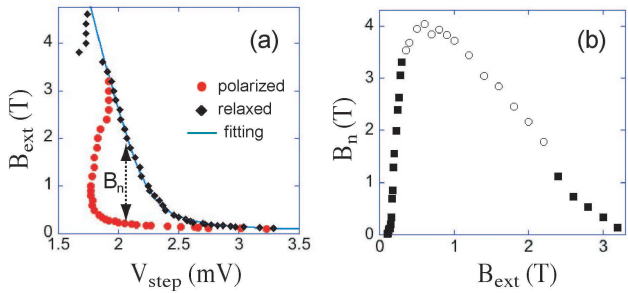}}
\caption{\label{fig:fig3} (a) Measured current step positions $V_{step}$ versus external field, for the two measurement sequences described in the text.  The data from the `relaxed' sequence have been numerically fit to an empirical function.  The differences along the vertical axis between the `polarized' data points and this curve yield estimates of the Overhauser fields $B_n$, which are plotted in $b$. (b) The values that rely on the extrapolated portion of the fitting curve in $a$ (e.g. for $B_{ext}>3.6$ T) are shown as open circles, and those that can be determined directly from the data points are shown as filled squares.  Experimental uncertainty is indicated by the size of the symbols. }
\end{figure}
\indent The experiments of  figure~\ref{fig:fig2} were repeated over a range of $B_{ext}$ values from $0.1-3.2$ T.  The results are summarized in figure~\ref{fig:fig3}.  The points in figure~\ref{fig:fig3}a are the measured values of $V_{step}$ versus $B_{ext}$ for the two measurement sequences described above.  The estimated Overhauser fields $B_n$ correspond to the differences between the two curves along the $B_{ext}$ axis.  The $B_n$ estimates plotted in figure~\ref{fig:fig3}b were obtained by subtracting the points from the `polarized' sequence from an empirical curve fit to the `relaxed' data points.  Note that figure~\ref{fig:fig3}b differentiates between $B_n$ values that can be determined directly from measured data points and those determined from the extrapolated portion of the fitted curve at $B_{ext}>3.6$ T in figure~\ref{fig:fig3}a.  The behavior observed in this latter range is most likely due to the polarization dynamics occurring near the zero detuning point, i.e. where the states $\ket{T_-}$ and $\ket{T_+}$ can both be nearly degenerate with the upper and lower singlet branches, respectively.  In figure~\ref{fig:fig3}b, $B_n$ rises rapidly in the range $0.1$T$<B_{ext}<0.4$T reaching its maximum value $\approx4$ T near $B_{ext}=0.65$ T.  It then falls off quasi-linearly to $\approx0$ at $B_{ext}=3.2$ T (notably, the sum $B_n+B_{ext}$ is roughly constant in this range $0.65$T$<B_{ext}<3.2$T).  \\
\indent Relaxation measurements of the quantity $B_n(\tau)$ were obtained by inserting a time delay $\tau$ at $V_{sd}=0$ after pumping in the `polarized' measurement sequence.  We observe a bi-exponential decay curve with approximate decay times of $13\pm3$ and $68\pm10$ seconds \footnote{The relaxation data was obtained at $B_{ext}=0.2$ T and did not change significantly at $B_{ext}=0.3$ T, however, we have not yet performed this measurement at other fields.}.  In general, this decay will be due to both diffusion of nuclear polarization away from the dot region and the intrinsic nuclear relaxation processes; however we expect intrinsic relaxation to be much slower than the rate of loss due to spin diffusion, and we estimate the latter to yield a $\sim$10-20s timescale (i.e. for diffusion in the vertical direction).\\
\indent Given the finite duration of the readout voltage sweep in this scheme, the true Overhauser fields should be slightly larger than these values, by a factor $\sim$1.2 for the largest fields, if a correction is made based on the measured relaxation.  Using the expression for $B_n$ in GaAs obtained in Ref.~\cite{Paget}, inserting $|g^*|=0.25$ from measurement of a nominally identical device \footnote{S. M. Huang and K. Ono, private communication.}, and assuming equal isotopic polarizations, we obtain an average nuclear polarization $P\approx B_n/(9.32 T)$.  The data therefore suggest a maximum polarization lying within the range $0.38-0.52$. \\
\begin{figure}
\scalebox{0.14}{\includegraphics{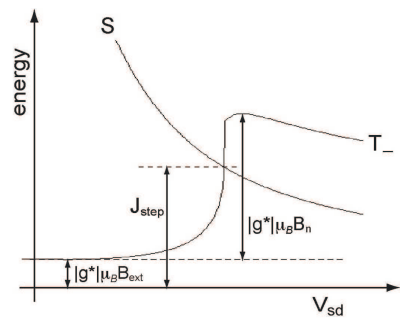}}
\caption{\label{fig:fig4} A schematic diagram showing the dynamics of the $\ket{T_-}$ energy level as the source-drain voltage $V_{sd}$ is swept from left to right and dynamic nuclear polarization occurs near the $E_{S/T_-}=0$ degeneracy.  $J_{step}$ indicates the value of $J$ at the crossing point, which is approximately the position (in $V_{sd}$) of the observed current step.}
\end{figure}
\indent To model this behavior, we consider the rate for an electron to escape the spin blockade due to hyperfine-induced singlet-triplet mixing.  The two-electron spin Hamiltonian $\ham=\ham_0+\ham'$ consists of the Zeeman and exchange terms ($\ham_0$) and perturbing hyperfine coupling terms ($\ham'$).  The triplet and singlet eigenstates of $\ham_0$ are mixed only by the hyperfine terms antisymmetric with respect to the two dots, i.e. the terms involving differences of their effective nuclear fields.  Let us define the antisymmetric nuclear fields $\tilde{h}_\alpha=\frac{A}{2}(\langle I^1_\alpha\rangle-\langle I^2_\alpha\rangle)$, where $\langle I^1_\alpha\rangle$ is the average $\alpha$ component of spin per nucleus in dot 1 (here we assume uniform hyperfine couplings for simplicity).  Statistical fluctuations give rise to fields $|\tilde{h}_\alpha|\sim A/\sqrt{N}$ where $N$ is the average number of nuclear spins.  The rate of nuclear spin $\Delta m_z=-1$ transitions ($\uparrow\rightarrow\downarrow$) is determined by the rate of escape from $\ket{T_-}$ via hyperfine mixing with $\ket{S(1,1)}$, and subsequent tunneling to the (0,2) and (0,1) charge configurations.  For $E_{S/T_-}>>\tilde{h}_{x,y}$, this can be treated by second-order perturbation theory \cite{Erlingsson} and the rate written as $W_{\uparrow\rightarrow\downarrow}=\rho_{T'_-}|\bra{S(1,1)}\ket{T'_-}|^2\Gamma_{S(1,1)\rightarrow(0,1)}$, where $\rho_{T'_-}$ is the steady-state occupation probability for the perturbed eigenstate $\ket{T'_-}$ and $\Gamma_{S(1,1)\rightarrow(0,1)}$ is the rate of escape from $\ket{S(1,1)}$ to (0,1) via (0,2).  In general, $\Gamma_{S(1,1)\rightarrow(0,1)}$ can reflect both elastic and inelastic processes, and therefore depends on tunnel couplings, phonon scattering rates and cotunneling processes; here it is only considered phenomenologically as a parameter.  First-order perturbation theory gives $\bra{S(1,1)}\ket{T'_-}=\frac{\tilde{h}_x+i\tilde{h}_y}{\sqrt{2}E_{S/T_-}}\sim\frac{A}{\sqrt{N}E_{S/T_-}}$.  Therefore $W_{\uparrow\rightarrow\downarrow}\approx\alpha(\frac{A/\sqrt{N}}{E_{S/T_-}})^2$, where $\alpha=\rho_{T'_-}\Gamma_{S(1,1)\rightarrow(0,1)}$ (the opposite rate $W_{\downarrow\rightarrow\uparrow}$ is computed the same way using the $\ket{T'_+}$ state). \\
\indent The rate of change of nuclear polarization in the $j^{th}$ dot can be expressed: 
\be
\label{eqn:dI/dt}
\frac{d\langle I^{j}_Z\rangle}{dt}=\frac{W_{\downarrow\rightarrow\uparrow}(1-p^{j}_{3/2})-W_{\uparrow\rightarrow\downarrow}(1-p^{j}_{-3/2})}{2N_{j}}-\frac{|\langle I^{j}_Z\rangle|}{T_1}
\ee 
where $p_{\pm3/2}$ are the normalized populations of the $m_z=\pm3/2$ sublevels and we assume an equal probability $1/2$ for a nuclear spin transition to occur in dot 1 or dot 2.  $T_1$ is the phenomenological relaxation rate of nuclear polarization (including spin diffusion out of the dots) and $N_j$ is the number of nuclear spins in the $j^{th}$ dot. As $V_{sd}$ is swept up, $E_{S/T_-}$ decreases (see figure~\ref{fig:fig1}) and $W_{\uparrow\rightarrow\downarrow}$ increases; a small nuclear polarization begins to accumulate.  Since $E_{S/T_-}=J-|g^*|\mu_B B_{ext}+A\langle I_Z\rangle$, $E_{S/T_-}$ is further decreased by the nuclear Overhauser field $A\langle I_Z\rangle=-|g^*|\mu_B B_{n}<0$ and positive feedback occurs.  Once this occurs, the system is forced to pass through the degeneracy $E_{S/T_-}=0$ where the pumping rate is maximum and is limited by the (non-nuclear spin-flip) escape rates of the $\ket{T_+}$ and $\ket{T_0}$ states \footnote{The leakage current $\sim$1 pA throughout the spin-blockade suggests a timescale $\sim$100 ns for these processes.  They enter the pumping rate $W_{\uparrow\rightarrow\downarrow}$ through the occupation probability $\rho_{T'_-}$.}.  This is illustrated schematically in figure~\ref{fig:fig4}.  A steady-state is then reached in the regime of $E_{S/T_-}<0$, at which the steady-state nuclear polarization can be determined by setting $\frac{d\langle I_Z\rangle_{ss}}{dt}=0$ and solving self-consistently for $\langle I_Z\rangle_{ss}$.  Assuming $W_{\uparrow\rightarrow\downarrow}>>W_{\downarrow\rightarrow\uparrow}$ and setting $p_{\pm3/2}=1/4$ (low polarization limit), we find for the steady-state polarization:  
\be
\label{eqn:pol}
P_{ss}\approx-\frac{2}{3}\sqrt[3]{\frac{3\alpha T_1}{8 N^2 (\beta-1)^2}}
\ee
where $\beta\equiv\frac{J-|g^*|\mu_B B_{ext}}{|g^*|\mu_B B^{ss}_{n}}$ is a free parameter with respect to this analytical solution but would be determined by a full simulation of the dynamics.  We take equation~\ref{eqn:pol} to be valid for $0<\beta\lesssim1/2$ \footnote{For $W_{\uparrow\rightarrow\downarrow}$ that is symmetric with respect to $E_{S/T_-}$, as we expect, $\beta$ should be $\approx 1/2$. At lower temperatures $W_{\uparrow\rightarrow\downarrow}$ could be dominated by asymmetric phonon emission and absorption rates, in which case it would be faster for $E_{S/T_-}<0$ (where the spin-flip is an energy relaxation process) and then $0<\beta\lesssim1/2$. In any case, $\beta=1$ is unphysical, so equation~\ref{eqn:pol} will not diverge.}.  Inserting $T_1=30$s, $N=5\times10^5$ and $\beta=1/2$, equation~\ref{eqn:pol} requires that $\alpha\sim 1.5\times10^8$ Hz for $|P_{ss}|\sim20\%$.  Such a fast rate for $\alpha$ (on the order of the dot-lead tunneling rate) is possible near the singlet-singlet resonance where $\ket{S(1,1)}$ and $\ket{S(0,2)}$ are hybridized by a relatively large interdot tunnel coupling $\sim50\mu$eV.  A full explanation of the dependence of $B_n$ as a function of $B_{ext}$, and of the current step \footnote{The steady-state rate $W^{ss}_{\uparrow\rightarrow\downarrow}\sim 10^3-10^4$ Hz is not large enough to explain the observed current step.  A reasonable explanation is that the two dots are not identical in size ($\Delta N=N_1-N_2\neq0$), and therefore an imbalance of nuclear polarization occurs according to equations~\ref{eqn:dI/dt} and~\ref{eqn:pol}. This imbalance causes an antisymmetric Overhauser field $\tilde{h}_z\approx\frac{2}{3}\frac{\Delta N}{N}A\langle I_Z\rangle$, which can be much larger than the fluctuating field, and generates additional current via mixing $\ket{T_0}\rightarrow\ket{S(1,1)}$.}, is beyond the scope of this paper, and will require a detailed modeling of the double-dot transport and polarization dynamics. However, we do expect $B_n$ to fall off as $B_{ext}$ is increased (as in figure~\ref{fig:fig3}b for $B_{ext}>1$T) due to increasing pumping of the opposite transition $W_{\downarrow\rightarrow\uparrow}$ via the state $\ket{T_+}$ mixing with the lower singlet branch.\\
\indent In conclusion,  we have shown that surprisingly large nuclear polarizations can be electrically induced and detected in the present GaAs double dot system.  One drawback is the detection scheme: the magnitude of polarization that can be detected is limited by the range of $B_{ext}$ over which we can observe a clear current step, and additionally by the fact that dot-lead tunneling rates will be reduced at large $B_{ext}$ due to a momentum mismatch between dot and lead states \cite{Zaslavsky}. Indeed, it is likely that significantly larger polarization was actually achieved in our experiments than was detected, since the current step observed in the readout $V_{sd}$ sweep suggests a secondary polarization boost.  Another detection method, such as electron spin resonance, could be used to extend these measurements in similar devices. \\
\indent We thank W. A. Coish, D. G. Austing and T. Kodera for stimulating discussions and A. Oiwa for experimental assistance. We acknowledge financial support from Grant-in-Aid for Scientific Research S (No. 19104007), SORST Interacting Carrier Electronics, JST, Special Coordination Funds for Promoting Science and Technology, and MEXT. J. B. acknowledges support through NSERC, CIAR, PREA, and a JSPS Fellowship. 
\bibliography{biblio}
\end{document}